\documentclass[12pt]{article}
\usepackage{epsf}
\usepackage[dvips]{graphicx,psfrag}

\renewcommand{\thefootnote}{\#\arabic{footnote}}
\begin{document}

\newcommand{\gtrsim}{ \mathop{}_{\textstyle \sim}^{\textstyle >} }
\newcommand{\lesssim}{ \mathop{}_{\textstyle \sim}^{\textstyle <} }

\renewcommand{\thefootnote}{\fnsymbol{footnote}}
\setcounter{footnote}{0}
\begin{titlepage}

\def\thefootnote{\fnsymbol{footnote}}

\begin{center}

\hfill TU-684\\
\hfill hep-ph/0304126\\
\hfill April, 2003\\

\vskip .5in

{\Large \bf

Cosmic String from $D$-term Inflation
and Curvaton

}

\vskip .45in

{\large
Motoi Endo$^{(a)}$, Masahiro Kawasaki$^{(b)}$ 
and Takeo Moroi$^{(a)}$
}

\vskip .3in

{\em
$^{(a)}$Department of Physics,  Tohoku University\\
Sendai 980-8578, Japan
}

\vskip .2in

{\em
$^{(b)}$Research Center for the Early Universe, School of Science,
University of Tokyo\\
Tokyo 113-0033, Japan
}

\end{center}

\vskip .4in

\begin{abstract}

We study effects of the cosmic string in the $D$-term inflation model
on the cosmic microwave background (CMB) anisotropy.  In the $D$-term
inflation model, gauged cosmic string is usually formed, which may
significantly affect the CMB anisotropy.  We see that its study
imposes important constraint.  In order to realize the minimal model
of $D$-term inflation, we see that the coupling constant $\lambda$ in
the superpotential generating the inflaton potential should be
significantly small, $\lambda\lesssim O(10^{-4}-10^{-5})$.  We also
discuss that a consistent scenario of the $D$-term inflation with
$\lambda\sim 1$ can be constructed by adopting the curvaton mechanism
where the cosmic density fluctuations are generated by a primordial
fluctuation of a late-decaying scalar field other than the inflaton.

\end{abstract}

\end{titlepage}

\renewcommand{\thepage}{\arabic{page}}
\setcounter{page}{1}
\renewcommand{\thefootnote}{\#\arabic{footnote}}
\setcounter{footnote}{0}

Inflation \cite{PRD23-347,MNRAS195-467} provides elegant solutions to
many of serious problems in cosmology.  As is well known, assuming the
de Sitter expansion of the universe in an early epoch, horizon and
flatness problems are solved since, during the inflation,
physical scale grows much faster than the horizon scale.  In addition,
quantum fluctuation during the inflation is now a very promising
candidate of the origin of the cosmic density fluctuations.  

From a particle-physics point of view, thus, it is important to find
natural and consistent models of inflation; indeed, many models of
inflation have been proposed so far.  It is, however, non-trivial to
construct a realistic model of inflation.  This is because the
potential of the inflaton, which is the scalar field responsible for
the inflation, is usually required to be very flat although, in
general, it is non-trivial to find a natural mechanism to guarantee
the flatness of the scalar potential.

Primarily, flatness of the scalar potential is disturbed by loop
effects.  In particular, quadratic divergences usually show up once
one calculates radiative corrections to the scalar mass and hence
scalar fields lighter than the cut-off scale (which is naturally of
the order of the gravitational scale) is hard to realize in a general
framework.  This difficulty can be relatively easily avoided by
introducing supersymmtry which we assume in this paper.  Even in the
supersymmetric framework, however, it is still difficult to find a
realistic model of inflation since, in the de Sitter background,
supergravity effects may induce effective mass as large as the
expansion rate during the inflation (i.e., so called ``Hubble-induced
mass'') to scalar fields, which spoils the flatness of the inflaton
potential.

One elegant solution to the problem of the Hubble-induced mass is to
adopt the $D$-term inflation \cite{PLB387-43,PLB388-241}.  Since the
Hubble-induced mass originates from $F$-term interactions, inflaton
potential is free from this problem if the vacuum energy during the
inflation is provided only by a $D$-term interaction, which is the
case in the $D$-term inflation model.

In the $D$-term inflation, a U(1) gauge interaction is introduced to
realize a flat potential of the inflaton.  Inflation proceeds in the
symmetric phase of the U(1) while, after the inflation, the U(1)
symmetry is broken by a vacuum expectation value of one of scalar
fields in the model.  Since the U(1) gauge symmetry is spontaneously
broken, gauged string arises in this framework and hence the $D$-term
inflation inevitably predicts the existence of the cosmic string
\cite{PRD56-6205,PLB412-28}.  If the cosmic string is formed, it
affects the CMB anisotropy and changes the shape of the CMB angular
power spectrum $C_l$ \cite{NATURE310-391,APJ288-422}, which is defined
as
\begin{eqnarray}
    \left\langle \Delta T(\vec{x}, \vec{\gamma}) 
        \Delta T(\vec{x}, \vec{\gamma}')  \right\rangle_{\vec{x}}
    = \frac{1}{4\pi} 
    \sum_l (2l+1) C_l P_l (\vec{\gamma} \cdot \vec{\gamma}'),
\end{eqnarray}
with $\Delta T (\vec{x}, \vec{\gamma})$ being the temperature
fluctuation of the CMB radiation pointing to the direction
$\vec{\gamma}$ at the position $\vec{x}$ and $P_l$ being the Legendre
polynomial.  If the effect of the cosmic string on the CMB anisotropy
is too strong, the CMB angular power spectrum may become inconsistent
with the observations since the observed values of $C_l$ is highly
consistent with that from purely adiabatic density fluctuations which
is the prediction of the ``usual'' inflation models without cosmic
string.

In this paper, we study effects of the cosmic string in the framework
of the $D$-term inflation paying particular attention to its effects
on the CMB anisotropy.  Taking account of the effect of the cosmic
string, we see that the CMB angular power spectrum may become
significantly deviate from the usual adiabatic results.  Using the
recent precise measurement of the CMB angular power spectrum by the
WMAP \cite{aph0302217}, we derive constraints on the $D$-term
inflation scenario.  As a result, we see that the coupling constant
$\lambda$ in the superpotential generating the inflaton potential
should be $O(10^{-4}-10^{-5})$ or smaller.  We also discuss that a
consistent scenario of the $D$-term inflation with $\lambda\sim 1$ is
possible by adopting the curvaton mechanism
\cite{PLB524-5,PLB522-215,PRD66-063501,PRD67-023503} where primordial
amplitude fluctuation of a late-decaying scalar condensation
(so-called curvaton) becomes the dominant origin of the cosmic density
fluctuations.\footnote
{For the generation of the cosmic density fluctuation from an
axion-like field in the pre-big-bang
\cite{PLB265-287,APP1-1,PRD50-2519} and the ekpyrotic
\cite{PRD64-123522,PRD65-086007,PRD66-046005} scenarios, see also
\cite{NPB626-395}.}

Let us start our discussion with a brief review of the $D$-term
inflation.  In the $D$-term inflation models, a new U(1) gauge
interaction is introduced, with three chiral superfields, $S(0)$,
$\bar{Q}(-1)$ and $Q(+1)$ where we denote the U(1) charges of the
superfields in the parenthesis.  With the superpotential
\begin{eqnarray}
W = \lambda S \bar{Q} Q,
\end{eqnarray}
and adopting non-vanishing Fayet-Illiopoulos $D$-term parameter $\xi$,
the scalar potential is given by
\begin{eqnarray}
V = \lambda^2 \left( |S\bar{Q}|^2 + |SQ|^2 + |\bar{Q}Q|^2 \right)
+ \frac{1}{2} g^2 \left(
-|\bar{Q}|^2 + |Q|^2 - \xi \right)^2,
\end{eqnarray}
where $g$ is the gauge coupling constant.\footnote
{Here and hereafter, we use the same notation for the scalar fields
and for the chiral superfields since there should be no confusion.}
In our study, we take $\xi$ to be positive (although the final result
is independent of this assumption).  Minimizing the potential, the
true vacuum is given by
\begin{eqnarray}
\langle S \rangle = 0,~~~
\langle \bar{Q} \rangle = 0,~~~
\langle Q \rangle = \sqrt{\xi}.
\label{truevac}
\end{eqnarray}

Although the true vacuum is given by (\ref{truevac}), there is a
(quasi) flat direction, that is, $S\rightarrow\infty$ with $\bar{Q}$
and $Q$ being vanished.  Indeed, in this limit, the scalar potential
becomes $V=\frac{1}{2}g^2\xi^2$ and hence, at the tree level, the
scalar potential becomes constant.  This flat direction is used as the
inflaton.

Once the radiative corrections are taken into account, the flat
direction is slightly lifted.  When the scalar field $S$ takes large
amplitude, $\bar{Q}$ and $Q$ become massive and decouple from the
effective theory at the energy scale $\lambda S$.  This fact means
that the gauge coupling constant in this case should be evaluated at
the scale $\lambda S$ and hence, for $S\gg g\sqrt{\xi}/\lambda$,
$V(S)=\frac{1}{2}g^2(\lambda S)\xi^2$.  Using one-loop renormalization
group equation, and defining
\begin{eqnarray}
S = \frac{1}{\sqrt{2}} \sigma e^{i\theta},
\end{eqnarray}
the potential for the real scalar field $\sigma$ is given by
\begin{eqnarray}
V(\sigma) = \frac{g^2}{2} \xi^2 +
\frac{g^4\xi^2}{8\pi^2} \log \frac{\sigma}{\sigma_0},
\end{eqnarray}
where $\sigma_0$ is some constant.  

Since the scalar field $\sigma$ has a very flat potential when
$\sigma$ is large, the $\sigma$ field can be used as an inflaton;
inflation occurs if $\sigma$ has large enough amplitude.  Assuming the
slow-roll condition, evolution of the $\sigma$ field during the
inflation is described as
\begin{eqnarray}
\sigma^2 = \sigma_{\rm end}^2 + \frac{g^2}{2\pi^2} N_e M_*^2,
\label{sigma(Ne)}
\end{eqnarray}
where $N_e$ is the $e$-folds of the inflation and $M_*\simeq 2.4\times
10^{18}\ {\rm GeV}$ is the reduced Planck scale.  The cosmic density
fluctuations responsible for the CMB anisotropy measured by the WMAP
(and other) experiment are generated when $N_e\sim 30-50$.
(Hereafter, we take $N_e=50$ in evaluating $C_l$.)  In addition,
$\sigma_{\rm end}$ is the inflaton amplitude at the end of the
inflation.  In order to realize the $D$-term inflation, $\sigma$
should be large enough so that (i) the slow-roll condition is
satisfied and (ii) the effective mass squared of the $Q$ field becomes
positive.  Inflation ends one of these conditions are violated and
hence $\sigma_{\rm end}$ is estimated as
\begin{eqnarray}
\sigma_{\rm end} = {\rm max}
(\sigma_{\rm s.r.}, \sigma_{\rm inst}),
\end{eqnarray}
where
\begin{eqnarray}
\sigma_{\rm s.r.}\simeq \frac{g}{2\pi}M_*,~~~
\sigma_{\rm inst}\simeq \frac{\sqrt{2}g}{\lambda} \sqrt{\xi}.
\end{eqnarray}
Notice that $\sigma_{\rm s.r.}$ and $\sigma_{\rm inst}$ are derived
from the slow-roll condition and the instability of the potential of
$Q$, respectively.

Once the evolution of the inflaton field is understood, we can
calculate the metric perturbation $\Psi$ generated from the primordial
fluctuation of the inflaton field.\footnote
{We adopt the notation used in \cite{HuPhD}.  For example, the
perturbed line element in the Newtonian gauge is given by
$ds^2=-(1+2\Psi)dt^2+a^2(1+2\Phi)\delta_{ij}dx^i dx^j $ with $a$ being
the scale factor.}
The metric perturbation after the inflation is proportional to the
following quantity \cite{PRD28-629}:
\begin{eqnarray}
{\cal R}^{\rm (inf)} \equiv
\frac{H_{\rm inf}}{2\pi} 
\frac{3H_{\rm inf}^2}{(\partial V/\partial\sigma)}
=
\sqrt{\frac{2}{3}}\pi \frac{\xi}{gM_*^3} \sigma,
\end{eqnarray}
where $H_{\rm inf}$ is the expansion rate during the
inflation.\footnote
{The spectral index $n_{\rm S}$ is very close to $1$ in the $D$-term
inflation and hence we neglect the scale dependence of the primordial
metric perturbation.}
(Here and hereafter, the superscript ``(inf)'' implies that the
quantity is induced from the fluctuation of the inflaton field.)
Importantly, ${\cal R}^{\rm (inf)}$ changes its behavior at
$\lambda\sim\lambda_{\rm crit}$ with
\begin{eqnarray}
\lambda_{\rm crit} \equiv 
\frac{2\pi}{\sqrt{N_e}} \frac{\sqrt{\xi}}{M_*}.
\end{eqnarray}
When $\lambda\gg\lambda_{\rm crit}$, the second term in the right-hand
side of Eq.\ (\ref{sigma(Ne)}) dominates over the first term and hence
$\sigma$ for corresponding $N_e$ is given by
$gM_*\sqrt{N_e}/\sqrt{2}\pi$.  On the contrary, if
$\lambda\ll\lambda_{\rm crit}$, the first term wins and
$\sigma\simeq\sigma_{\rm inst}$.  As a result, we obtain
\begin{eqnarray}
{\cal R}^{\rm (inf)}
\simeq \frac{1}{\sqrt{3}} 
\frac{\xi}{M_*^2}\sqrt{N_e} \times
\left\{
\begin{array}{ll}
1 & ~~~:~~~ \lambda\gg\lambda_{\rm crit}
\\
(\lambda/\lambda_{\rm crit})^{-1}
& ~~~:~~~ \lambda\ll\lambda_{\rm crit}
\end{array}
\right. .
\label{Psi(inf)}
\end{eqnarray}
When $\lambda\gg\lambda_{\rm crit}$, ${\cal R}^{\rm (inf)}$ is
proportional to $\xi$ and is independent of $\lambda$.  On the
contrary, if $\lambda\ll\lambda_{\rm crit}$, ${\cal R}^{\rm (inf)}$ is
proportional to $\xi^{3/2}/\lambda$.  This fact implies that, for a
fixed value of $\xi$, the metric perturbation generated from the
inflaton fluctuation is enhanced for sufficiently small value of
$\lambda$.

Once the non-vanishing metric perturbation is generated, it becomes an
origin of the cosmic density fluctuations.  One important point is
that the density fluctuations associated with ${\cal R}^{\rm (inf)}$
are purely adiabatic.  Notice that the metric perturbation depends on
time and, for superhorizon modes, the metric perturbation in the
radiation dominated epoch is given by
\begin{eqnarray}
\Psi^{\rm (inf)}_{\rm RD} = \frac{2}{3} 
{\cal R}^{\rm (inf)}.
\end{eqnarray}
(Here and hereafter, the subscript ``RD'' is for quantities in the
radiation dominated epoch.)

After the inflation, scalars settle to the values given in
(\ref{truevac}).  Thus, the gauged U(1) symmetry is spontaneously
broken and the cosmic string is formed.  The mass per unit length of
the string is given by
\begin{eqnarray}
\mu = 2 \pi \xi.
\label{mu=2pixi}
\end{eqnarray}

Once the cosmic-string network is formed, it affects the cosmic
density fluctuations.  In particular, an important constraint is
obtained by studying its effects on the CMB anisotropy.  The
perturbations induced by cosmic strings are non-Gaussian and
decoherent isocurvature, which leads to characteristic spectrum of the
CMB anisotropy, which is distinguished from that induced by inflation.

The CMB angular power spectrum in the $D$-term inflation scenario
contains two contributions: one is the adiabatic one from the
primordial fluctuation of the inflaton and the other is from the
cosmic string.  Assuming no correlation between these two
contributions, we obtain
\begin{eqnarray}
C_l = C_l^{\rm (inf)} + C_l^{\rm (str)},
\end{eqnarray}
where $C_l^{\rm (inf)}$ and $C_l^{\rm (str)}$ are contributions from
primordial inflaton fluctuation and cosmic string, respectively.  The
adiabatic part $C_l^{\rm (inf)}$ can be calculated with the
conventional method; we use the CMBFAST package \cite{APJ469-437} to
calculate $C_l^{\rm (inf)}$.  (In our study, we use the cosmological
parameters $\Omega_bh^2=0.024$, $\Omega_mh^2=0.14$, $h=0.72$, and
$\tau=0.166$, where $\Omega_b$ and $\Omega_m$ are density parameters
of baryon and non-relativistic matter, respectively, $h$ the Hubble
constant in units of 100\ km/sec/Mpc, and $\tau$ the optical depth,
which are suggested from the WMAP experiment \cite{aph0302209}.  It
should be noted that these may not necessarily be the best-fit values
in the $D$-term inflation case.  Varying of these parameters within
the reasonable ranges, however, does not change our main conclusion.)
Since $C_l^{\rm (inf)}$ is from the two point correlation function,
the CMB angular power spectrum is second order in density $\Psi^{\rm
(inf)}$.  In particular, $C_l^{\rm (inf)}$ is proportional to $\xi^2$
and $\xi^3/\lambda^2$ for $\lambda\gg\lambda_{\rm crit}$ and
$\lambda\ll\lambda_{\rm crit}$, respectively.  In addition, the cosmic
string contribution $C_l^{\rm (str)}$ is proportional to
$\mu^2\propto\xi^2$ \cite{VilShe}.

For quantitative studies on the effects of the cosmic string,
$l$-dependence of $C_l^{\rm (str)}$ should be known.  Usually
numerical methods are used for this purpose but, unfortunately, it is
rather difficult to determine the detailed shape of $C_l^{\rm (str)}$;
there are two classes of results which give different $l$-dependence
of $C_l^{\rm (str)}$.  One class of results show relatively ``flat''
behavior of the CMB angular power spectrum
\cite{PRL77-3061,PRL79-2624,aph0302166}; that is, $l(l+1)C_l^{\rm
(str)}$ is found to be almost constant (at least up to the multipole
$l\sim 300$ \cite{PRL79-2624}).  However, another class of studies
result in ``tilted'' behavior
\cite{PRL82-679,PRD59-023508,PRD60-083504,PRD61-063505}; the function
$\sqrt{l(l+1)C_l^{\rm (str)}}$ approximately has a linear dependence
on $\log l$ up to $l\sim 400-600$ then it steeply decreases.  There is
still some discussion on this issue and it has not been clearly
understood how $C_l^{\rm (str)}$ behaves.

Thus, in our study, we consider both cases adopting some approximate
formulae of $C_l^{\rm (str)}$.  For the flat case, shape of $C_l^{\rm
(str)}$ at high multipole is quite uncertain.  Thus, assuming a
dumping behavior at multipole higher than $l_{\rm max}$, we
parameterize the cosmic-string contribution as
\begin{eqnarray}
l(l+1)C_l^{\rm (str)} &=& 
\left\{ \begin{array}{ll}
A^{\rm (str)}_{\rm SW} T_{\rm CMB}^2 (G\mu)^2 
&~~~:~~~ l \leq l_{\rm max}
\\
0 
&~~~:~~~ l > l_{\rm max}
\end{array} \right. .
\label{Cl^str(flat)}
\end{eqnarray}
Numerical calculation suggests $l_{\rm max}\gtrsim 300$
\cite{PRL79-2624}.  On the contrary, for the tilted case, we use the
formula
\begin{eqnarray}
l(l+1)C_l^{\rm (str)} &=& 
A^{\rm (str)}_{\rm SW} T_{\rm CMB}^2 (G\mu)^2 
\nonumber \\ &&
\times
\left\{ \begin{array}{ll}
\left[ 1 + a_{l<l_{\rm kink}} \log (l/10) \right]^2
&~~~:~~~ l \leq l_{\rm kink}
\\
\left[ 1 + a_{l<l_{\rm kink}} \log (l_{\rm kink}/10) 
+ a_{l>l_{\rm kink}} \log(l/l_{\rm kink}) \right]^2
&
~~~:~~~ l > l_{\rm kink}
\end{array} \right. ,
\nonumber \\
\label{Cl^str(tilted)}
\end{eqnarray}
where $l_{\rm kink}$ is the multipole above which $C_l^{\rm (str)}$ is
suppressed for the tilted case.  Numerical results suggest $l_{\rm
kink}\sim 400-600$, and we use $l_{\rm kink}=500$ in our following
discussion.  In addition, hereafter, we take $a_{l>500}=-0.6$.  The
values of $A^{\rm (str)}_{\rm SW}$ and $a_{l<l_{\rm kink}}$ (for the
tilted case) found in the literatures are listed in Table
\ref{table:a&r}.\footnote
{For a discussion on the model-dependence of the numerical results,
see also \cite{PRD59-023508}.}
As one can see, although the numerical values for $A^{\rm (str)}_{\rm
SW}$ are fairly scattered, $A^{\rm (str)}_{\rm SW}$ is found to be
$\sim O(100)$.

\begin{table}[t]
\begin{center}
\begin{tabular}{ccc}
\hline\hline
{$A_{\rm str}$}
& {$a_{l<l_{\rm kink}}$}
& {Ref.} \\
\hline
{$130-350$}  & {$\sim 0$ (flat)}  
& {\cite{PRL77-3061,PRL79-2624}} \\
{$\sim 700$} & {$\sim 0$ (flat)}  
& {\cite{aph0302166}} \\
{$\sim 700$} & {$0.24-0.34$}  
& {\cite{PRL82-679}} \\
{$\sim 600$} & {$0.26-0.49$}
& {\cite{PRD60-083504}} \\
{$\sim 500$} & {$0.18-0.23$}  
& {\cite{PRD61-063505}} \\
\hline\hline
\end{tabular}
\caption{Values of $A^{\rm (str)}_{\rm SW}$ and $a_{l<l_{\rm kink}}$
found in the literatures.  (See Eqs.\ (\ref{Cl^str(flat)}) and
(\ref{Cl^str(tilted)}).)}
\label{table:a&r}
\end{center}
\end{table}

With Eqs.\ (\ref{Cl^str(flat)}) and (\ref{Cl^str(tilted)}), effect of
the cosmic string on the Sachs-Wolfe (SW) tail is given by
\begin{eqnarray}
[l(l+1)C_l^{\rm (str)}]_{l=10} \simeq
2.2\times 10^{3} {\rm \mu K}^2 \times
\left(\frac{\sqrt{\xi}}{10^{16}{\rm GeV}}\right)^4
\left(\frac{A^{\rm (str)}_{\rm SW}}{100}\right),
\end{eqnarray}
while the inflaton contribution is found to be
\begin{eqnarray}
[l(l+1)C_l^{\rm (inf)}]_{l=10} \simeq
2.1\times 10^{3} {\rm \mu K}^2 \times
\left(\frac{\sqrt{\xi}}{10^{16} {\rm GeV}}\right)^4
\times {\rm max}[1, (\lambda/\lambda_{\rm crit})^{-2}].
\end{eqnarray}
Thus, the CMB angular power spectrum is expected to be significantly
affected by the cosmic string contribution (at least at the SW tail)
assuming $A^{\rm (str)}_{\rm SW}\gtrsim O(100)$, unless
$\lambda\ll\lambda_{\rm crit}$.  This fact implies that the parameter
$\lambda$ plays a significant role in determining the shape of the
total angular power spectrum.  For a given value of $\xi$, $\Psi^{\rm
(inf)}$ is more enhanced as $\lambda$ decreases.  Thus, when
$\lambda\ll\lambda_{\rm crit}$, $C_l$ can be dominated by the inflaton
contribution and hence it (almost) agrees with the adiabatic result.
If $\lambda$ is larger than $\lambda_{\rm crit}$, on the contrary,
$C_l^{\rm (inf)}$ and $C_l^{\rm (str)}$ are both proportional to
$\xi^2$ and are independent of $\lambda$.  Thus, relative size of the
inflaton and cosmic-string contributions are fixed in this case.

The CMB angular power spectrum measured by the WMAP is well explained
by the adiabatic result \cite{aph0302217}.  Thus, based on the fact
that $C_l^{\rm (inf)}$ and $C_l^{\rm (str)}$ have different
$l$-dependence, the cosmic-string contribution should not be too
large.  Defining 
\begin{eqnarray}
\tilde{A}^{\rm (str)}_{\rm SW} \equiv
\left[ 
\frac{\Psi^{\rm (inf)}}
{\Psi^{\rm (inf)}|_{\lambda\gg\lambda_{\rm crit}}}
\right]^{-2}
A^{\rm (str)}_{\rm SW},
\end{eqnarray}
the total angular power spectrum is determined by this quantity up to
the normalization since $C_l^{\rm (str)}\propto A^{\rm (str)}_{\rm
SW}\mu^2\propto A^{\rm (str)}_{\rm SW}[\Psi^{\rm
(inf)}|_{\lambda\gg\lambda_{\rm crit}}]^2$ while $C_l^{\rm
(inf)}\propto[\Psi^{\rm (inf)}]^2$.  Notice that $\tilde{A}^{\rm
(str)}_{\rm SW}\simeq A^{\rm (str)}_{\rm SW}$ for
$\lambda\gtrsim\lambda_{\rm crit}$ while $\tilde{A}^{\rm (str)}_{\rm
SW}\simeq (\lambda/\lambda_{\rm crit})^2A^{\rm (str)}_{\rm SW}$ for
$\lambda\lesssim\lambda_{\rm crit}$.

\begin{figure}[t]
    \begin{center}
        \scalebox{0.75}{\includegraphics{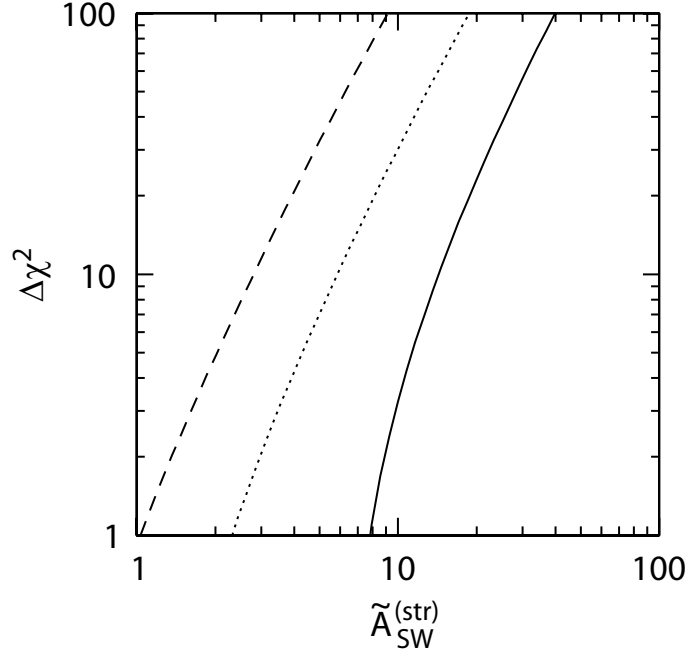}}
        \caption{The $\chi^2$ variable as a function of $A_{\rm str}$
        for the flat case with $l_{\rm max}=300$ (solid) and tilted
        case with $a_{l<500}=0.2$ (dotted) and $0.4$ (dashed).}
        \label{fig:chi2astr}
    \end{center}
\end{figure} 

With the approximated formulae (\ref{Cl^str(flat)}) and
(\ref{Cl^str(tilted)}), we calculate the $\chi^2$ variable as a
function of $\tilde{A}^{\rm (str)}_{\rm SW}$ and $\xi$ (or
equivalently, ${\cal R}^{\rm (inf)}$) using the likelihood code
provided by the WMAP collaboration \cite{aph0302218} with the WMAP
data \cite{aph0302217}.  For a fixed value of $\tilde{A}^{\rm
(str)}_{\rm SW}$, we vary $\xi$ and obtain the minimal value of
$\chi^2$, which we call $\chi^2_{\rm min}(\tilde{A}^{\rm (str)}_{\rm
SW})$.  We checked that the $\chi^2$ variable has its minimum value
when $\tilde{A}^{\rm (str)}_{\rm SW}$ is small enough, i.e., when the
cosmic-string contribution is negligible.  (When $\tilde{A}^{\rm
(str)}_{\rm SW}=0$, we found that the best-fit value of ${\cal R}^{\rm
(inf)}$ is given by $[{\cal R}^{\rm (inf)}]_{\rm best}\simeq 4.5\times
10^{-5}$ which gives $\chi^2_{\rm min}\simeq 975$.)  Then, we
calculate
\begin{eqnarray}
\Delta\chi^2 = 
\chi^2_{\rm min} (\tilde{A}^{\rm (str)}_{\rm SW}) - 
\chi^2_{\rm min} (\tilde{A}^{\rm (str)}_{\rm SW}=0).
\end{eqnarray}
The $\Delta\chi^2$ variable for several cases are plotted in Fig.\ 
\ref{fig:chi2astr}.  As one can see, $\Delta\chi^2_{\rm min}$
drastically increases once $\tilde{A}_{\rm str}$ becomes larger than
$O(1-10)$.

From $\Delta\chi^2$, we can determine the upper bound on
$\tilde{A}^{\rm (str)}_{\rm SW}$.  For $\Delta\chi^2<4$
($\Delta\chi^2<9$), upper bound on $\tilde{A}^{\rm (str)}_{\rm SW}$
for the flat case (with $l_{\rm max}=300$) and tilted case (with
$a_{l<500}=0.2$) is found to be
\begin{eqnarray}
\tilde{A}^{\rm (str)}_{\rm SW} \lesssim 
\left\{ \begin{array}{ll}
10.6~ (13.8) & ~~~:~~~{\rm flat~} (l_{\rm max}=300)
\\
4.0~ (5.6) & ~~~:~~~{\rm tilted~} (a_{l<500}=0.2)
\end{array} \right. .
\label{Astr_bound}
\end{eqnarray}
Since $\Delta\chi^2$ drastically increases once $\tilde{A}^{\rm
(str)}_{\rm SW}$ becomes larger than $\sim 10$, the above upper bound
does not become much larger than $10$ irrespective of the detailed
shape of $C_l^{\rm (str)}$.  For example, for the flat case,
$\Delta\chi^2<4$ results in $\tilde{A}^{\rm (str)}_{\rm SW}\lesssim
7.5$ ($5.1$, $9.1$) for $l_{\rm max}=200$ ($400$, $1000$).  Even for
$l_{\rm max}=20$, we obtain $\tilde{A}^{\rm (str)}_{\rm SW}\lesssim
13.9$.  In addition, for the tilted case, we obtain $\tilde{A}^{\rm
(str)}_{\rm SW}\lesssim 2.6$ and $1.9$ for $a_{l<500}=0.3$ and $0.4$,
respectively.  Thus, we conclude that $\tilde{A}^{\rm (str)}_{\rm
SW}\gg 10$ is highly inconsistent with the observation.

\begin{figure}[t]
    \begin{center}
        \scalebox{0.75}{\includegraphics{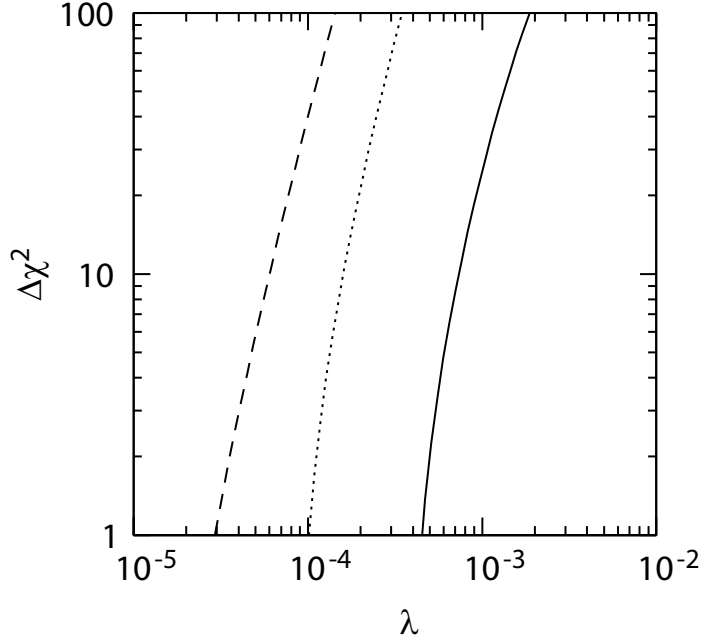}}
        \caption{The $\chi^2$ variable as a function of $\lambda$
        for the flat case with $l_{\rm max}=300$ and $A^{\rm
        (str)}_{\rm SW}=100$ (solid), $l_{\rm max}=300$ and $A^{\rm
        (str)}_{\rm SW}=700$ (dotted), and for the tilted case with
        $a_{l<500}=0.2$ and $A^{\rm (str)}_{\rm SW}=700$ (dashed).}
        \label{fig:chi2lam}
    \end{center}
\end{figure} 

Above upper bounds on $\tilde{A}^{\rm (str)}_{\rm SW}$ are much
smaller than the reasonable value of $A^{\rm (str)}_{\rm SW}$ (i.e.,
$A^{\rm (str)}_{\rm SW}\sim O(100)$).  In other words, if we adopt
typical value of $A^{\rm (str)}_{\rm SW}$ given in the literatures,
$\lambda$ should be sufficiently small to suppress the cosmic-string
contribution.  In order to derive upper bound on $\lambda$, we
calculate the $\chi^2$ variable for the flat and tilted cases; we vary
$\xi$ to minimize $\chi^2$ for fixed value of $\lambda$ and obtain
$\chi^2_{\rm min}(\lambda)$, the minimum value of $\chi^2$ for a fixed
value of $\lambda$.  In Fig.\ \ref{fig:chi2lam}, we plot
$\Delta\chi^2=\chi^2_{\rm min}(\lambda)-\chi^2_{\rm
min}(\lambda\ll\lambda_{\rm crit})$ as a function of $\lambda$.  As
one can see, $\lambda$ is required to be smaller than
$O(10^{-4}-10^{-5})$.  For $\Delta\chi^2<4$, the upper bound on
$\lambda$ is given by $5.8\times 10^{-4}$ ($1.3\times 10^{-4}$,
$4.5\times 10^{-5}$) for the flat case with $l_{\rm max}=300$ and
$A^{\rm (str)}_{\rm SW}=100$ (for the flat case with $l_{\rm max}=300$
and $A^{\rm (str)}_{\rm SW}=700$, for the tilted case with
$a_{l<500}=0.2$ and $A^{\rm (str)}_{\rm SW}=700$).  

If the cosmic-string contribution is negligibly small, the WMAP result
can be well described by the $D$-term inflation scenario with ${\cal
R}^{\rm (inf)}\simeq 4.5\times 10^{-5}$.  Using Eq.\ (\ref{Psi(inf)}),
we obtain the best-fit value of the $\xi$ as
\begin{eqnarray}
\sqrt{ [\xi]_{\rm best} } \simeq 2.6 \times 10^{15}\ {\rm GeV} \times 
\left( \frac{\lambda}{10^{-4}} \right)^{1/3}.
\end{eqnarray}

We would like to note here that a consistent scenario of the $D$-term
inflation with $\lambda\sim 1$ can be constructed by adopting the
curvaton mechanism
\cite{PLB524-5,PLB522-215,PRD66-063501,PRD67-023503}.  In the curvaton
scenario, the dominant part of the cosmic density fluctuations are
generated from the primordial fluctuation of a late-decaying scalar
field $\phi$ (other than inflaton), i.e., so called the curvaton
field.\footnote
{Another possibility may be to break the U(1) symmetry even during the
inflation \cite{PLB412-28}.  For this purpose, however, several new
fields with U(1) charge should be introduced.  They induces new
$D$-flat direction and hence the inflaton potential may be affected.
Breaking the U(1) symmetry during the inflation is therefore highly
non-trivial.}
The curvaton field $\phi$ has non-vanishing initial amplitude
$\phi_{\rm init}$.  In addition, the curvaton potential (during
inflation) is so flat that $\phi$ acquires quantum fluctuation during
inflation:
\begin{eqnarray}
\delta\phi_{\rm init}=\frac{H_{\rm inf}}{2\pi}.
\label{phi_init}
\end{eqnarray}
Just after the decay of the inflaton, the curvaton is assumed to be a
sub-dominant component of the universe, so the $\delta\phi_{\rm init}$
produces an entropy fluctuation in the curvaton sector.  As the
universe expands, however, the curvaton dominates the energy density
of the universe and then decays to reheat the universe.  Important
point is that the primordial fluctuation $\delta\phi_{\rm init}$ is
converted to purely adiabatic density fluctuations (as far as all the
components in the universe, i.e., the baryon, the cold dark matter,
and so on as well as the radiation are generated from the decay
product of the curvaton).  Since $H_{\rm inf}$ does not change much
during the $D$-term inflation, (almost) scale-invariant adiabatic
cosmic density fluctuations can be realized if the curvaton
contribution is large enough.  Effect of $\delta\phi_{\rm init}$ on
the CMB anisotropy is parameterized by a single parameter, which we
choose to be the metric perturbation (in the RD epoch after the decay
of the curvaton field) induced by the curvaton fluctuation
\cite{PLB522-215,PRD66-063501}:
\begin{eqnarray}
\Psi^{\rm (curv)}_{\rm RD} = 
-\frac{4}{9} \frac{\delta\phi_{\rm init}}{\phi_{\rm init}}
= -\frac{2}{9\sqrt{6}\pi} \frac{g\xi}{M_*\phi_{\rm init}}.
\end{eqnarray}

Since there is no correlation between the primordial fluctuations in
the inflaton and curvaton fields, the total CMB angular power spectrum
is now given in the form
\begin{eqnarray}
C_l = C_l^{\rm (inf)} + C_l^{\rm (str)} + C_l^{\rm (curv)},
\label{C_l(w/curvaton)}
\end{eqnarray}
where $C_l^{\rm (curv)}$ is the curvaton contribution.  As we
emphasized, the density fluctuations associated with $\delta\phi_{\rm
init}$ is purely adiabatic.  In addition, $\Psi^{\rm (curv)}$ is
almost scale-invariant since the expansion rate during the $D$-term
inflation is almost constant.  As a result, the total CMB angular
power spectrum may become well consistent with the WMAP observation if
the curvaton contribution dominates over the inflaton contribution.
We can see that such a hierarchy is relatively easily realized; using
the fact that $C_l^{\rm (inf)}$ and $C_l^{\rm (str)}$ are proportional
to $[\Psi^{\rm (inf)}]^2\propto\mu^2$ when $\lambda\sim 1$ while
$C_l^{\rm (curv)}$ is proportional to $[\Psi^{\rm (curv)}]^2$, the
curvaton contribution dominates when $\phi_{\rm init}\lesssim gM_*$
since $\Psi^{\rm (inf)}\sim O(\xi/M_*^2)$.

To be more quantitative, we calculate the $\chi^2$ variable to derive
constraints on the parameters in the scenario.  Now the total angular
power spectrum has the form
\begin{eqnarray}
C_l = [C_l^{\rm (adi)}]_{\Psi_{\rm RD}=1} \Psi_{\rm RD}^2 
+ [C_l^{\rm (str)}]_{G\mu=1} (G\mu)^2,
\end{eqnarray}
where
\begin{eqnarray}
\Psi_{\rm RD}^2 = 
[\Psi^{\rm (inf)}_{\rm RD}]^2+[\Psi^{\rm (curv)}_{\rm RD}]^2,
\end{eqnarray}
and $C_l^{\rm (adi)}$ is the CMB angular power spectrum generated from
purely adiabatic density fluctuations.  Then, as we discussed in the
case without the curvaton, the $\chi^2$ variable is minimized when the
cosmic-string contribution is negligibly small and $\Psi_{\rm
RD}\simeq 3.0\times 10^{-5}$.  The latter condition can be satisfied
by tuning the curvaton contribution (as far as the inflaton
contribution to $\Psi_{\rm RD}$ is not too large.)  In order to
realize hierarchy between the adiabatic and cosmic-string
contributions, it is necessary to make $\mu$ small by suppressing
$\xi$.  Requiring that $\Delta\chi^2<4$, we obtain the upper bound on
$\mu$ for the case of $\lambda\sim 1$ as
\begin{eqnarray}
A^{\rm (str)}_{\rm SW}  \times (G\mu/10^{-6})^2 
\lesssim
\left\{ \begin{array}{ll}
80 & ~~~:~~~{\rm flat~} (l_{\rm max}=300)
\\
30 & ~~~:~~~{\rm tilted~} (a_{l<500}=0.2)
\end{array} \right. ,
\end{eqnarray}
and hence, using Eq.\ (\ref{mu=2pixi}), the upper bound on $\xi$ is
given by
\begin{eqnarray}
\sqrt{\xi} / M_* \lesssim 
(A^{\rm (str)}_{\rm SW}/100)^{-1/4} \times
\left\{ \begin{array}{ll}
1.9 \times 10^{-3} & ~~~:~~~{\rm flat~} (l_{\rm max}=300)
\\
1.5 \times 10^{-3} & ~~~:~~~{\rm tilted~} (a_{l<500}=0.2)
\end{array} \right. .
\end{eqnarray}
In addition, using the best-fit value of $\Psi_{\rm RD}$ given above,
we can estimate the required value of the initial amplitude of the
curvaton field.  Importantly, in order to realize the adiabatic-like
CMB angular power spectrum, $\Psi^{\rm (curv)}\gg\Psi^{\rm (inf)}$
since $C_l^{\rm (str)}$ and $C_l^{\rm (inf)}$ are of the same order if
$\lambda\sim 1$.  Thus, requiring $\Psi^{\rm (curv)}_{\rm RD}\simeq
3.0\times10^{-5}$, we obtain
\begin{eqnarray}
\phi_{\rm init}/M_* \simeq 
9.5 \times 10^{2}
\times g \left( \frac{\sqrt{\xi}}{M_*} \right)^2.
\end{eqnarray}

We should also note here that, in the $D$-term inflation scenario,
Hubble-induced mass term of the scalar fields does not exist.  This
fact has an important implication to the curvaton scenario.  The
curvaton field should acquire sizable primordial quantum fluctuation.
Thus, the curvaton potential should be sufficiently flat during
inflation, which is naturally realized in the $D$-term inflation.  In
models where the vacuum energy during the inflation is from $F$-term
interaction, however, it is non-trivial to realize the flatness of the
curvaton potential during the inflation.  Thus, the $D$-term inflation
provides a consistent scenario of the curvaton paradigm.

Finally, we would like to make a brief comment on the cases of hybrid
inflation.  In many classes of hybrid inflation models, gauged U(1)
symmetry exists which is spontaneously broken after the inflation.
With such a U(1) symmetry, cosmic string is also formed, which also
affects the CMB anisotropy. The dynamics of the hybrid inflation and
the mass of the string per unit length are almost the same as the
D-term inflation \cite{PRD61-083512}. Therefore, the result obtained
in the present work can apply to the hybrid inflation.  Of course, for
general hybrid inflation models where the cosmic string affects the
CMB anisotropy too much to be consistent with the observations, the
curvaton mechanism can solve the difficulty as in the $D$-term
inflation case.

\end{document}